\documentclass[aip,jcp,preprint]{revtex4} 
\usepackage{graphicx} 
\usepackage{amsmath} 
\usepackage{amssymb} 

\begin{document} 
\title{Polarizable Surfaces: Weak and Strong Coupling Regimes} 

\author{Alexandre P. dos Santos} 
\email{alexandre.pereira@ufrgs.br} 
\affiliation{Instituto de F\'isica, Universidade Federal do Rio Grande do Sul, Caixa Postal 15051, CEP 91501-970, Porto Alegre, RS, Brazil} 

\author{Yan Levin} 
\email{levin@if.ufrgs.br} 
\affiliation{Instituto de F\'isica, Universidade Federal do Rio Grande do Sul, Caixa Postal 15051, CEP 91501-970, Porto Alegre, RS, Brazil} 

\begin{abstract}
We study the ionic distribution near a charged surface.  A new method for performing Monte Carlo simulations in this geometry is discussed. A theory is then presented  that allows us to accurately reproduce the density profiles obtained in the simulations.  In  the weak-coupling regime, a theory accounts for the ion-image interactions, leading to a modified Poisson-Boltzmann equation. When the correlations between the ions are significant, a strong-coupling theory is used to calculate the density profiles near the surface and a Poisson-Boltzmann equation with a renormalized boundary condition to account for the counterion distribution in the far-field. 
\end{abstract}

\maketitle 

\section{Introduction}\label{intro}
Study of charged surfaces in electrolyte solutions is of fundamental importance, 
since these can model lamellar liquid crystals, clays, biological membranes, electrodes, etc. Interesting phenomena such as like-charge attraction between similarly charged surfaces
has been observed in the presence of 
multivalent counterions~\cite{GuJo84,PeCa97,Le02}. There has been a great 
theoretical~\cite{EnWe78,KjMi97,Ne01,MoNe01,LaPi02,JhPa07,AbAn07,JhKa08,HaLu09,HaLu10,SaTr11a,SaTr11b}, simulational~\cite{GuJo84,PeCa97,MoNe02,WaMa12}, and experimental~\cite{DuLe04} effort to  clarifying the behavior of double layers near charged surfaces. In many approaches the theories assume that the entire system is composed of the same dielectric material.
This, however, is not very realistic since clays, colloidal particles, and hydrocarbon membranes, have dielectric constant significantly smaller than that of the surrounding aqueous medium. The dielectric discontinuity across the interface results in polarization effects~\cite{JhPa07,JhKa08,HaLu09,HaLu10,DoBa11,LuLi11,WaMa12,GaXi12} which can significantly affect the ionic
distribution near the surface. In the present chapter, we present a
simple theoretical approach which allows us to accurately predict the counterion distribution near a charged wall which separates two environments with different dielectric constants. We consider separately the weak and the strong coupling regimes. Monte Carlo simulations are also performed in order to test our theoretical predictions.

\section{Monte Carlo Simulations}\label{mcs}

The simulations of long-range interacting systems are much more difficult than of systems with short-range forces.  The difficulty is that one can
not arbitrarily cut off the long-range Coulomb potential by using periodic boundary conditions, as is the case of the usual Lennard-Jones fluids. Instead one needs to consider an infinite number of periodic images of the system and then sum over these using Ewald summation methods~\cite{AlTi87}. For systems with a planar geometry, such as an infinite charged wall in contact with an electrolyte, there is an additional complication which comes from the broken translational symmetry. In this section, we describe an approach that allows us to simulate such systems taking into account the dielectric discontinuity at the interface. The Monte Carlo~(MC) simulations are performed in the NVT ensemble. The system is located in the right-hand half of a rectangular simulation box of dimensions $L_{x} \times L_{y} \times L_{z}$, centered at the origin of coordinate system. A charged wall of surface charge density $-\sigma$ is 
located at $z=0$. $N_c=\text{int}\left[\sigma L_{xy}^2/q\alpha\right]$ neutralizing counterions 
of charge $\alpha q$ and effective radius $r_c$, are confined to the region $0<z<L_z/2$, where $q$ is the proton charge and $\alpha$ is the ionic valence. The dielectric constants on the two sides of the wall are different, given by $\epsilon_c$ and $\epsilon_w$, for $z<0$ and $z>0$, respectively. Note that the dielectric discontinuity results in the appearance of the image charges  in the region $-L_z/2 < z < 0$, which will be discussed later. The  Ewald  summation~\cite{AlTi87} is used in order to calculate the electrostatic potentials between the ions in the periodic replicas of the simulation box. To account for the slab geometry we use the correction proposed by Yeh and Berkowitz~\cite{YeBe99}. The complete derivation of the electrostatic energy is presented in the appendix.

\section{Theory: Weak Regime}

We first present a theory that accounts for the results of the MC simulations in the weak coupling limit, when the characteristic Coulomb interaction between the counterions is smaller than the thermal energy, $\Gamma \equiv \alpha^2 q^2/\epsilon_w d k_B T \ll 1$,
where $d$ is the characteristic distance between the condensed counterions. 
Using $\alpha q/\pi d^2=\sigma$, the plasma parameter becomes $\Gamma=\sqrt{\alpha^3 q^3 \pi \sigma }/\epsilon_w k_B T$. The Bjerrum length is defined as $\lambda_B=\beta q^2/\epsilon_w$ and is $7.2$~\AA, for water at room temperature.

\begin{figure}[h]
\begin{center}
\includegraphics[width=8cm]{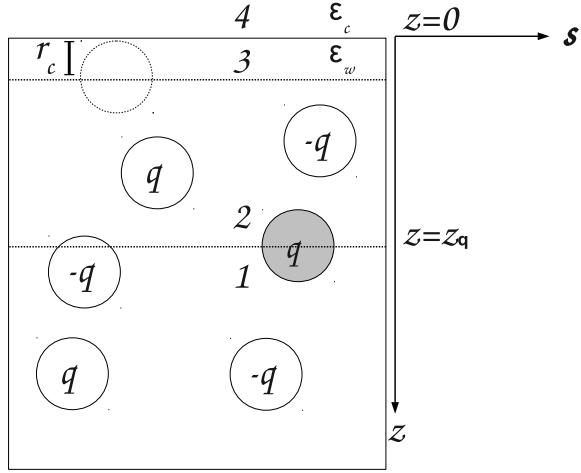}
\end{center}
\caption{Representation of an electrolyte in the region $z>r_c$.}
\label{fig1}
\end{figure}

Before studying the ionic distribution near a charged wall, we first need to
understand the role of electrostatic correlations and the induced charges when
$\sigma=0$.  To this end we consider a symmetric $\alpha$:$\alpha$ electrolyte at
concentration $c_s$ confined to infinite half-space, Fig.~\ref{fig1}. The work necessary to bring an ion from the bulk to a distance $z_q$ from the (uncharged) surface which separates the two regions with the different dielectric constants, $\epsilon_w$ and $\epsilon_c$, can be calculated in terms of the electrostatic Green's function~\cite{LeFl01}.


To account for the interionic correlations and induced surface charge, we use the linearized Poisson-Boltzmann (Debye-H\"uckel) equation. For symmetry reasons it is convenient to work in cylindrical coordinate system. 
Suppose that an ion of charge $q$ is located at $z_q$, see Fig.~\ref{fig1}. The electrostatic potential inside the regions $1$ and $2$ satisfies 
\begin{equation}
\nabla^2\phi({\pmb s},z)-\kappa^2 \phi({\pmb s},z)=-\frac{4 \pi \alpha q}{\epsilon_w} \delta({\pmb s})\delta(z-z_q)\ , 
\end{equation}
while in the regions $3$ and $4$ it satisfies the Laplace equation,
\begin{equation}
\nabla^2\phi({\pmb s},z)=0 \ ,
\end{equation}
where $\kappa=\sqrt{8\pi \alpha^2 \lambda_B c_s}$ is the inverse Debye length.

Writing the potential as a Fourier transform, $\phi({\pmb s},z)=(1/4\pi^2)\int_{-\infty}^{+\infty}d{\pmb k}\ e^{i{\pmb k}\cdot{\pmb s}} \ \hat\phi({\pmb k},z)$, we obtain the following equation for the regions $1$ and $2$,
\begin{equation}\label{dh}
\frac{\partial^2\hat\phi({\pmb k},z)}{\partial z^2}-(k^2+\kappa^2) \hat\phi({\pmb k},z)= -\frac{4 \pi \alpha q}{\epsilon_w} \delta(z-z_q)\ ,
\end{equation}
and for the regions $3$ and $4$,
\begin{equation}
\frac{\partial^2\hat\phi({\pmb k},z)}{\partial z^2}=k^2 \hat\phi({\pmb k},z) \ ,
\end{equation}
where we have used a Fourier representation of the delta function 
\begin{equation}
\delta({\pmb s})=\frac{1}{(2\pi)^2}\int_{-\infty}^{+\infty}d{\pmb k}\ e^{i{\pmb k}\cdot{\pmb s}} \ .
\end{equation}
Since the electrostatic potential must remain finite in the limits $z\rightarrow \infty$ and $z\rightarrow -\infty$, we obtain the following solutions for each region:
\begin{eqnarray}
\begin{array}{l}
\hat\phi_1({\pmb k},z)=B_1 e^{-pz} \ , \\
\hat\phi_2({\pmb k},z)=A_2 e^{pz}+B_2 e^{-pz} \ , \\
\hat\phi_3({\pmb k},z)=A_3 e^{kz}+B_3 e^{-kz} \ , \\
\hat\phi_4({\pmb k},z)=A_4 e^{kz} \ ,
\end{array}
\end{eqnarray}
where $p=\sqrt{k^2+\kappa^2}$. 

To calculate the integration constants, we use the conditions of continuity of the electrostatic potential, $\hat\phi_3({\pmb k},z)=\hat\phi_4({\pmb k},z)$ at $z=0$, $\hat\phi_2({\pmb k},z)=\hat\phi_3({\pmb k},z)$ at $z=r_c$ and $\hat\phi_2({\pmb k},z)=\hat\phi_1({\pmb k},z)$ at $z=z_q$, and  of the normal components of the displacement field,
\begin{eqnarray}
\begin{array}{l}
\epsilon_c \frac{\partial \hat\phi_4({\pmb k},z) }{\partial z}-\epsilon_w \frac{\partial \hat\phi_3({\pmb k},z) }{\partial z}=0 \text{ , at  } z=0 \ , \\ \epsilon_w \frac{\partial \hat\phi_3({\pmb k},z) }{\partial z}-\epsilon_w \frac{\partial \hat\phi_2({\pmb k},z) }{\partial z}=0 \text{ , at  } z=r_c \ , \\ \epsilon_w \frac{\partial \hat\phi_2({\pmb k},z) }{\partial z}-\epsilon_w \frac{\partial \hat\phi_1({\pmb k},z) }{\partial z}=4\pi \alpha q \text{ , at  } z=z_q \ .
\end{array}
\end{eqnarray}
The last equation has been obtained by integrating Eq.~\ref{dh} across the singularity
at $z_q$.

The Fourier transform of the electrostatic potential in the region $2$ is found to be
\begin{equation}\label{hatphi1}
\hat\phi_2({\pmb k},z)= \frac{2\pi \alpha q}{\epsilon_w p} \left[ e^{-p(z_q-z)} + e^{-p(z+z_q-2 r_c)}\frac{f_1(k)}{f_2(k)} \right] \ ,
\end{equation}
where
\begin{eqnarray}
f_1(k)=p \cosh{(k r_c)}-k \sinh{(k r_c)}+\frac{\epsilon_c}{\epsilon_w}p\sinh{(k r_c)}-\nonumber \\
\frac{\epsilon_c}{\epsilon_w}k \cosh{(k r_c)} \ ,
\end{eqnarray}
\begin{eqnarray}
f_2(k)=p \cosh{(k r_c)}+k \sinh{(k r_c)} + \frac{\epsilon_c}{\epsilon_w}p\sinh{(k r_c)}+\nonumber \\
\frac{\epsilon_c}{\epsilon_w}k \cosh{(k r_c)}
\end{eqnarray}
and the inverse Fourier transform is
\begin{equation}\label{hatphi2}
\phi_2({\pmb s},z)=\frac{1}{2\pi}\int_{0}^{\infty}dk\ k J_0(ks)\hat\phi_2({\pmb k},z) \ ,
\end{equation}
where $J_0(ks)$ is the Bessel function of order $0$. 

We are interested in calculating the potential felt by an ion, located at distance
$z_q$ from the interface. Subtracting the self-potential $q/\epsilon_w(z_q-z)$,
after performing the explicit integration of the first term in Eq.~\ref{hatphi1},  
we find
\begin{equation}
\phi_{pol}(z_q)= -\frac{\alpha q \kappa}{\epsilon_w} + \frac{\alpha q}{\epsilon_w} \int_0^\infty dk \  e^{-2p(z_q- r_c)}\frac{k\ f_1(k)}{p\ f_2(k)}  \ .
\end{equation}
Performing the G\"untelberg charging process~\cite{Gu26}, we obtain the work necessary to bring an ion from the bulk to a distance $z_q$ from the interface~\cite{LeFl01},
\begin{equation}\label{image_potential_i}
W_i(z_q)=\frac{\alpha^2 q^2}{2\epsilon_w} \int_0^\infty dk \  e^{-2p(z_q- r_c)}\frac{k\ f_1(k)}{p\ f_2(k)}  \ .
\end{equation}
A very accurate approximation to the above expression is
\begin{equation}\label{wzap}
W_{ap}(z_q)=\frac{W_i(r_c) r_c}{z_q} \ e^{-2 \kappa (z_q-r_c)} \ .
\end{equation}
This approximate form is much more convenient for numerical implementation~\cite{LeDo09,DoDi10b},
since it requires calculating only one integral to determine  $W_i(r_c)$ 
at the beginning of the calculation.

\begin{figure}[h]
\begin{center}
\includegraphics[width=8cm]{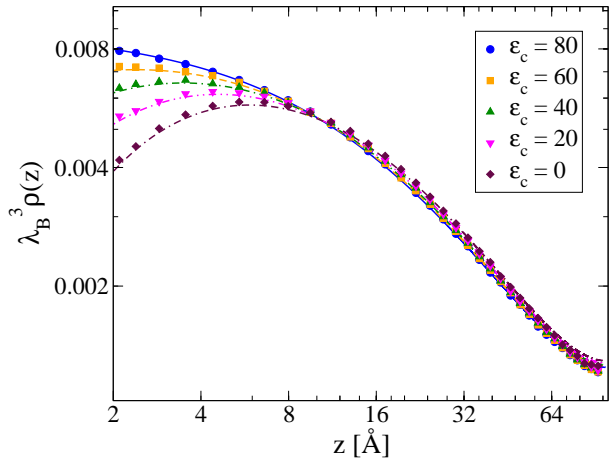}
\end{center}
\caption{The symbols are the simulation data, while the lines represent the solutions of the modified PB equation (Eq.~\ref{pbnoi}). The surface charge density is $\sigma = 6.25\times 10^{-4}~q/$\AA$^2$ and the monovalent counterion radius is $r_c=2$~\AA.}
\label{fig2}
\end{figure}

We now return to the problem of interest. The system now is an infinite dielectric wall of charge density $-\sigma$, located at $z=0$, and the neutralizing counterions of charge 
$\alpha q$ and radius $r_c$, confined to $0<z<L_z/2$. The dielectric constants are
$\epsilon_c$ and $\epsilon_w$, for $z<0$ and $z>0$, respectively. For $\Gamma<1$ (weak coupling limit) the electrostatic potential and the ionic density profile can be determined from the solutions of the modified PB equation
\begin{equation}\label{pbnoi}
\nabla^2 \phi(z) = -\frac{4\pi}{\epsilon_w}\left[-\sigma \delta(z) + \alpha q \rho(z)  \right] \ , 
\end{equation}
where the counterion density is given by
\begin{equation}\label{charge_dens}
\rho(z) = \frac{\sigma\ e^{-\alpha q\beta\phi(z)-\beta W_{ap}(z)}}{\alpha q \ \int_{r_c}^{L_z/2} dz \ e^{-\alpha q\beta\phi(z)-\beta W_{ap}(z)}}  \ . 
\end{equation}
The ionic correlations and the surface polarization are taken into account through the potential $W_{ap}(z)$, with $\kappa=\sqrt{8\pi \lambda_B \alpha \sigma/q L_z}$. In  Fig.~\ref{fig2}, we compare our results with the MC simulations, for various dielectric constants. As can be seen, the agreement between the theory and the simulations is excellent.

\section{Strong Coupling Regime}

\begin{figure}[h]
\begin{center}
\includegraphics[width=8cm]{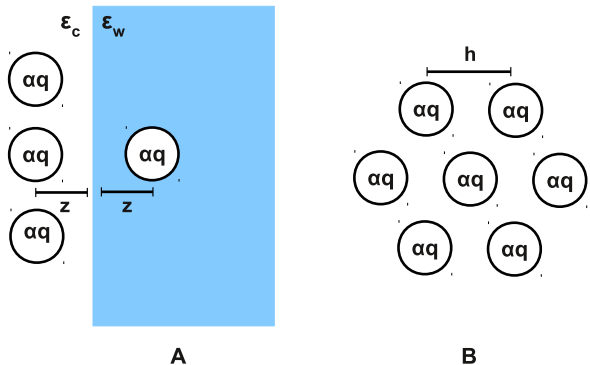}
\end{center}
\caption{Hexagon of images at the surface. In (A) the side view. In (B) the self image and the nearest neighbors. In order to illustrate we consider $\epsilon_c=0$.}
\label{fig3}
\end{figure}

When $\Gamma>1$, the mean field theory --- such as the PB equation --- is not able to accurately predict the ionic density distribtuion, because of the strong correlations between the counterions. In the limit $\Gamma \gg 1$, the counterions form a quasi-two dimensional strongly correlated liquid near the wall~\cite{Sh99,Le02}, with an approximately hexagonal geometry~\cite{To75}. Consider one counterion. The electric fields produced by the others counterions of the double layer approximately cancel each other. The counterion then interacts predominantly with the wall and with the ionic image charges, see Fig.~\ref{fig3}. The potential produced by the charged plate which separates the two environments with different dielectric constants is given by
\begin{equation}\label{phi_p}
\phi_p(z)=-\frac{4 \pi \sigma}{(\epsilon_w+\epsilon_c)} z \ . 
\end{equation}
As an approximation, we consider that the ion interacts only with the self-image  and with the image charges of the 6 first neighbors in the hexagonal lattice, see Fig.~\ref{fig3}.

This approximation was used previously in the study of colloidal double layers~\cite{BaDo11}. The electrostatic energy  of a counterion at distance $z$ from the plate is then
\begin{equation}\label{total}
U(z)=\alpha q \phi_p(z) + \frac{\gamma\alpha^2 q^2}{\epsilon_w 4 z}+\frac{6\gamma\alpha^2 q^2}{\epsilon_w \sqrt{4z^2+h^2}} \ ,
\end{equation}
where $\gamma=(\epsilon_w-\epsilon_c)/(\epsilon_w+\epsilon_c)$ and $h$ is the distance between the ions of the hexagonal lattice. $h$ can be calculated by considering that $N_c=\sigma A/\alpha q$ ions are distributed on the surface of
area $A$. The unitary cell of a hexagonal lattice is a parallelogram of area $h^2\sqrt{3}/2$, which gives the result
\begin{equation}
h=\sqrt{\dfrac{2\alpha q}{\sigma \sqrt{3}}} \ .
\end{equation}

\begin{figure}[h]
\begin{center}
\includegraphics[width=8cm]{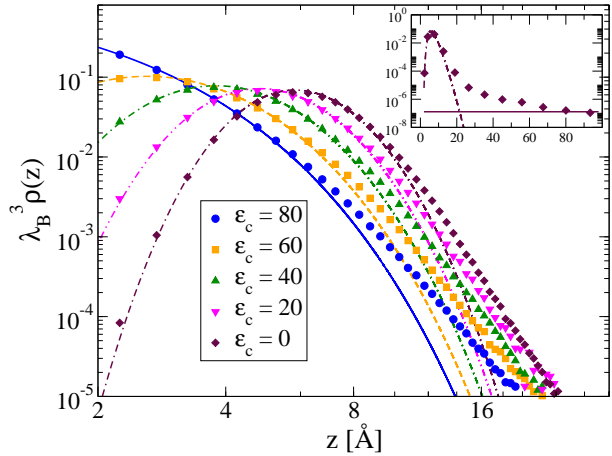}
\end{center}
\caption{The symbols are simulation data, while the lines represent the theory. The surface charge density is $\sigma = 3.74\times 10^{-3}~q/$\AA$^2$ and the pentavalent counterions radius is $r_c=2$~\AA. The solid line in the inset shows the solution of the regular PB equation with the boundary condition given by Eq.~\ref{nbc}.}
\label{fig4}
\end{figure}

The ionic density profile near the surface is obtained from 
\begin{equation}\label{bol}
\rho(z)= C e^{-\beta U(z)}
\end{equation} 
where $C=\sigma/\alpha q\int_0^L dz\ e^{-\beta U(z)} $ is the normalization constant. 
In Fig.~\ref{fig4} we compare our theoretical results with the MC simulations. The agreement is very good in the region where the strong coupling approximation applies.

In the far field, we expect that the counterions will be very dilute so that the
electrostatic potential will, once again, satisfy the PB equation.  The boundary
condition at the colloidal surface, however, must be modified to account for the strong
counterion condensation induced by the electrostatic correlations.  The new boundary conditon can be derived by equating the electrochemical potential of the condensed 
counterions and of the counterions which remain in the bulk~\cite{Sh99,DoDi09,DoDi10a}.  
This results in a new boundary condition for the standard PB equation which requires that the
concentration of the counterions near the surface be
\begin{equation}\label{nbc}
\rho_{PB}(0)= \rho_{sc}e^{\beta\mu_c} \ ,
\end{equation}
where $\beta \mu_c=-1.65\Gamma + 2.61\Gamma^{1/4} -0.26 \ln \Gamma -1.95$ is the chemical potential of the strongly correlated counterions~\cite{To75}. The density $\rho_{sc}$ is obtained using the coarse-graining of the near-field density profile, Eq.~\ref{bol}, in the region near the surface~\cite{DoDi09},
\begin{equation}
\rho_{sc}=\frac{\int_{r_c}^{r_c+3.6\lambda_{GC}} dz\ \rho(z)}{3.6\lambda_{GC}} \ ,
\end{equation}
where $\lambda_{GC}=1/2 \pi \alpha \lambda_B \sigma $ is the Gouy-Chapman length. In the inset of the Fig.~\ref{fig4} we present the solution of the usual PB equations with the renormalized boundary condition given by Eq.~\ref{nbc}. Only the case with $\epsilon_c=0$ is shown, since in the far field  the ionic density distribution is highly insensitive to the value of $\epsilon_c$.

\section{Conclusions}

We have presented a method for performing MC simulations in a cell geometry that includes
a dielectric discontinuity at one of the boundaries.  The results of the simulation
have been used to study the counterion density profiles and to develop the
weak and the strong coupling theories which account very accurately for the simulation data. In the weak coupling regime, the image charges repel the counterions from the wall. The contact density predicted by the present theory is substantially smaller than is found using the usual PB equation, and is in excellent agreement with the MC simulations. 
In the strong coupling limit,
the contact density is found to be even lower, since in this case the counterions are  repelled both by the self-image and by the images of the others counterions.  Finally, we
show how for $\Gamma \gg 1$ the counterion density distribution can be calculated in the far field using a renormalized boundary condition for the standard PB equation. 

In presenting the theory we have restricted ourselves to the systems containing only counterions and no coions.  In the weak coupling limit, the approach developed here 
can be easily extended to systems which also contain 1:1 electrolyte.  
The situation, however,
is much more difficult for multivalent electrolytes.  For such systems, strong
electrostatic interactions between the counterions and coions  lead to 
formation of Bjerrum clusters.  Thus, to be able to
account for the distribution of multivalent ions near a charged surface one
must first have an accurate description of the bulk of solution.  This, already presents a formidable challenge, see Ref.~\cite{Le02}.  Nevertheless, one can make 
some progress by considering
a chemical picture of  electrolyte in which there is an equilibrium between the free ions and the clusters, such calculations, however, very rapidly become quite 
involved~\cite{DoDi10a}.

\section{Acknowledgments}\index{acknowledgments}
This work was partially supported by the CNPq, FAPERGS, INCT-FCx, and by the US-AFOSR under the grant FA9550-09-1-0283.

\begin{appendix}\label{appe}
\section{Energy Calculation for Monte Carlo Simulations}

We consider a charge neutral system of $N$ ions of charges $q_i$. The electrostatic potential at the position ${\pmb r}$, created by all ions
(excluding ion $i$), their image charges (including the image of ion $i$), and the periodic replicas is
\begin{eqnarray}\label{elec_pot}
\phi_i({\pmb r})=\sum_{\pmb n}^{\infty} \sum_{j=1}^{N}{}^{'}\int\frac{\rho_j({\pmb s})}{\epsilon_w
|{\pmb r}-{\pmb s}+{\pmb r}_{ep}|}d^3{\pmb s} + \nonumber \\
\sum_{\pmb n}^{\infty} \sum_{j=1}^{N}\int\frac{\rho_j'({\pmb s})}{\epsilon_w
|{\pmb r}-{\pmb s}+{\pmb r}_{ep}|}d^3{\pmb s} \ ,
\end{eqnarray}
where $\rho_j({\pmb s})=q_j \delta({\pmb s}-{\pmb r}_j-{\pmb r}_{ep})$ and
$\rho_j'({\pmb s})=\gamma q_j \delta({\pmb s}-{\pmb r}'_j-{\pmb r}_{ep})$ are the
charge densities of  ions and their replicas;  and of dielectric images
and their replicas.  The replication vector is defined as ${\pmb
r}_{ep}=L_{xy}n_x\hat{\pmb x}+L_{xy}n_y\hat{\pmb y}+L_{z}n_z\hat{\pmb z}$ and ${\pmb r}'_j={\pmb r}_j-2z_j \hat{\pmb z}$. The
vectors ${\pmb n}=(n_x,n_y,n_z)$, where $n_x$, $n_y$ and $n_z$ are integers, 
represent the infinite replicas  of the main
cell. The constant $\gamma$ is
defined as $\gamma=(\epsilon_w-\epsilon_c)/(\epsilon_w+\epsilon_c)$ and the
prime on the summation means that $j\neq i$, when ${\pmb n}=(0,0,0)$. The total
electrostatic energy of the system is given by
\begin{equation}
U=\frac{1}{2}\sum_{i=1}^{N}q_i\phi_i({\pmb r}_i) \ .
\end{equation} 
The energy above is very difficult to calculate because of the slow
convergence of the series in Eq.~\ref{elec_pot}. To speed up the convergence, 
we use the Ewald method in which the ionic charge is 
partially screened by placing a Gaussian-distributed
charge of opposite sign on top of each ion~\cite{AlTi87}. We then add and subtract opposite Gaussian charge at the position of each ion 
and its image, $\rho_j({\pmb s})$ and $\rho'_j({\pmb s})$, respectively.
The potential, Eq.~\ref{elec_pot}, then becomes
\begin{equation}\label{elec_pot2}
\phi_i({\pmb r})=\phi_i^S({\pmb r})+\phi^L({\pmb r})-\phi_i^{self}({\pmb r}) \ ,
\end{equation}
where
\begin{eqnarray}\label{phi_S}
\phi_i^S({\pmb r})=\sum_{\pmb n}^{\infty}\sum_{j=1}^{N}{}^{'}\int\frac{\rho_j({\pmb s})-\rho_j^G({\pmb
s})}{\epsilon_w |{\pmb r}-{\pmb s}+{\pmb r}_{ep}|}d^3{\pmb s} + \nonumber \\
\sum_{\pmb n}^{\infty}\sum_{j=1}^{N}\int\frac{\rho_j'({\pmb s})-\rho_j'^{G}({\pmb s})}{\epsilon_w |{\pmb
r}-{\pmb s}+{\pmb r}_{ep}|}d^3{\pmb s} \ ,
\end{eqnarray}
\begin{eqnarray}\label{phi_L}
\phi^L({\pmb r})=\sum_{\pmb n}^{\infty}\sum_{j=1}^{N}\int\frac{\rho_j^G({\pmb s})}{\epsilon_w |{\pmb
r}-{\pmb s}+{\pmb r}_{ep}|}d^3{\pmb s} + \nonumber \\
\sum_{\pmb n}^{\infty}\sum_{j=1}^{N}\int\frac{\rho_j'^{G}({\pmb
s})}{\epsilon_w |{\pmb r}-{\pmb s}+{\pmb r}_{ep}|}d^3{\pmb s}
\end{eqnarray}
and
\begin{equation}\label{phi_self}
\phi^{self}_i({\pmb r})= \int\frac{\rho_i^G({\pmb s})}{\epsilon_w |{\pmb r}-{\pmb
s}|}d^3{\pmb s} \ ,
\end{equation}
where $\rho_j^G({\pmb
s})=q_j (\kappa_e^3/\sqrt{\pi^3})\exp{(-\kappa_e^2|{\pmb s}-{\pmb r}_j-{\pmb
r}_{ep}|^2)}$, $\rho_j'^G({\pmb s})=\gamma
q_j(\kappa_e^3/\sqrt{\pi^3})\exp{(-\kappa_e^2|{\pmb s}-{\pmb r}'_j-{\pmb
r}_{ep}|^2)}$ and $\kappa_e$ is a dumping parameter. We subtracted the self
potential, Eq.~\ref{phi_self}, from the Eq.~\ref{elec_pot2}, in order
to remove the prime over the summation in the long-range ($L$) 
part of the potential, Eq.~\ref{phi_L}. The electrostatic potential produced by the Gaussian charges
can be easily calculated using the Poisson equation, yielding
\begin{eqnarray}\label{phi_L2}
\phi^L({\pmb r})=\sum_{\pmb n}^{\infty}\sum_{j=1}^{N}
q_j\frac{\text{erf}{(\kappa_e |{\pmb r}-{\pmb r}_j+{\pmb r}_{ep}|)}}{\epsilon_w
|{\pmb r}-{\pmb r}_j+{\pmb r}_{ep}|} + \nonumber \\
\sum_{\pmb n}^{\infty}\sum_{j=1}^{N}\gamma q_j \frac{\text{erf}{(\kappa_e
|{\pmb r}-{\pmb r}'_j+{\pmb r}_{ep}|)}}{\epsilon_w |{\pmb r}-{\pmb r}'_j+{\pmb r}_{ep}|} \ ,
\end{eqnarray}
where $\text{erf}(x)$ is the error function. The short-range part of the 
potential ($S$), Eq.~\ref{phi_S}, can then be obtained 
in terms of the complementary error function,
$\text{erfc}(x)=1-\text{erf}(x)$,
\begin{eqnarray}\label{phi_S2}
\phi_i^S({\pmb r})=\sum_{\pmb n}^{\infty}\sum_{j=1}^{N}{}^{'}
q_j\frac{\text{erfc}{(\kappa_e |{\pmb r}-{\pmb r}_j+{\pmb r}_{ep}|)}}{\epsilon_w
|{\pmb r}-{\pmb r}_j+{\pmb r}_{ep}|} + \nonumber \\
\sum_{\pmb n}^{\infty}\sum_{j=1}^{N}\gamma q_j\frac{\text{erfc}{(\kappa_e
|{\pmb r}-{\pmb r}'_j+{\pmb r}_{ep}|)}}{\epsilon_w |{\pmb r}-{\pmb r}'_j+{\pmb r}_{ep}|} \ .
\end{eqnarray}
This potential decays very rapidly and can be truncated by setting the
dumping parameter to $\kappa_e=5/V^{1/3}$, where $V=L_{xy}^2L_z$, corresponding to the
minimum image convention. 
It is then sufficient to consider in the sum only the term 
${\pmb n}=(0,0,0)$,  with the usual periodic boundary condition, 
\begin{equation}\label{phi_S3}
\phi_i^S({\pmb r})=\sum_{j=1}^{N}{}^{'} q_j\frac{\text{erfc}{(\kappa_e |{\pmb
r}-{\pmb r}_j|)}}{\epsilon_w |{\pmb r}-{\pmb r}_j|} + \sum_{j=1}^{N}\gamma q_j
\frac{\text{erfc}{(\kappa_e |{\pmb r}-{\pmb r}'_j|)}}{\epsilon_w |{\pmb r}-{\pmb
r}'_j|} \ .
\end{equation}
The self-potential, Eq.~\ref{phi_self}, reduces to
\begin{equation}\label{phi_self2}
\phi_i^{self}({\pmb r})=q_i\frac{\text{erf}{(\kappa_e |{\pmb r}-{\pmb
r}_i|)}}{\epsilon_w |{\pmb r}-{\pmb r}_i|} \ .
\end{equation}
We next calculate the long-range part of the potential, Eq.~\ref{phi_L2}.
This is most easily obtained using the Fourier representation,  $\hat{\phi}^L({\pmb
k})=(1/V)\int_Vd^3{\pmb r}\ \exp{(-i{\pmb k}\cdot{\pmb r})}\phi^L({\pmb
r})$, since in the reciprocal space all the sums, once again, converge very rapidly. The Fourier transform $\hat{\rho}^T({\pmb
k})=(1/V)\int_Vd^3{\pmb r}\ \exp{(-i{\pmb k}\cdot{\pmb r})}\rho^T({\pmb r})$, of the Gaussian charge density,
\begin{eqnarray}
\rho^T({\pmb r})=\sum_{\pmb n}^{\infty}\sum_{j=1}^{N}
q_j\frac{\kappa_e^3}{\sqrt{\pi^3}}
\exp{(-\kappa_e^2|{\pmb r}-{\pmb r}_j-{\pmb r}_{ep}|^2)} + \nonumber \\
\sum_{\pmb n}^{\infty}\sum_{j=1}^{N} \gamma q_j\frac{\kappa_e^3}{\sqrt{\pi^3}}
\exp{(-\kappa_e^2|{\pmb r}-{\pmb r}'_j-{\pmb r}_{ep}|^2)} \ ,
\end{eqnarray}
is 
\begin{eqnarray}
\hat{\rho}^T({\pmb k})= \frac{1}{V} \exp{(-\frac{|{\pmb
k}|^2}{4\kappa_e^2})}\left[\sum_{j=1}^{N}q_j\exp{(-i{\pmb k}\cdot{\pmb r}_j)} \right.+ \nonumber \\
\left. \sum_{j=1}^{N}\gamma q_j\exp{(-i{\pmb k}\cdot{\pmb r}'_j)} \right] \ ,
\end{eqnarray}
where ${\pmb k}=(2\pi n_x/L_{xy},2\pi n_y/L_{xy},2\pi n_z/L_{z})$. Using
the Poisson equation, $|{\pmb k}|^2 \hat{\phi}^L({\pmb
k})=(4\pi/\epsilon_w) \hat{\rho}^T({\pmb k})$, we can evaluate the Fourier
transform of the potential,
\begin{eqnarray}
\hat{\phi}^L({\pmb k})=\frac{4\pi}{\epsilon_w V |{\pmb k}|^2}
\exp{(-\frac{|{\pmb k}|^2}{4\kappa_e^2})} \left[ \sum_{j=1}^{N}q_j\exp{(-i{\pmb
k}\cdot{\pmb r}_j)} + \right.\nonumber \\
\left.\sum_{j=1}^{N}\gamma q_j\exp{(-i{\pmb k}\cdot{\pmb r}'_j)} \right] \ .
\end{eqnarray}
The corresponding real-space  electrostatic potential is calculated using the inverse Fourier
transform, $\phi^L({\pmb r})=\sum_{{\pmb k}}\hat{\phi}^L({\pmb k})\exp{(i{\pmb k}\cdot{\pmb r})}$,
\begin{eqnarray}\label{phi_L3}
 \phi^L({\pmb r})=\sum_{{\pmb k}}\frac{4\pi}{\epsilon_w V |{\pmb k}|^2}
\exp{(-\frac{|{\pmb k}|^2}{4\kappa_e^2})}\exp{(i{\pmb k}\cdot{\pmb r})}\times \nonumber \\
\left[\sum_{j=1}^{N}q_j\exp{(-i{\pmb k}\cdot{\pmb r}_j)} + 
\sum_{j=1}^{N}\gamma q_j\exp{(-i{\pmb k}\cdot{\pmb r}'_j)} \right] \ .
\end{eqnarray}

The long-range contribution to the total electrostatic energy is given by
$U_L=(1/2)\sum_{i=1}^{N}q_i\phi^L({\pmb r}_i)$, where $\phi^L({\pmb
r})$ is obtained from Eq.~\ref{phi_L3}. It is convenient to rewrite this 
in terms of functions: $A({\pmb k})= \sum_{i=1}^{N}q_i \cos{({\pmb k}\cdot {\pmb
r}_i)}$, $B({\pmb k})=- \sum_{i=1}^{N}q_i \sin{({\pmb k}\cdot {\pmb r}_i)}$,
$C({\pmb k})= \sum_{i=1}^{N}\gamma q_i \cos{({\pmb k}\cdot {\pmb r}'_i)}$ and
$D({\pmb k})=- \sum_{i=1}^{N} \gamma q_i \sin{({\pmb k}\cdot {\pmb r}'_i)}$. The
electrostatic energy then becomes,
\begin{eqnarray}\label{U_long}
U_L = \sum_{{\pmb k}}\frac{2\pi}{\epsilon_w V |{\pmb k}|^2}
\text{exp}(-\frac{|{\pmb k}|^2}{4\kappa_e^2}) \times \nonumber \\
\left[A({\pmb k})^2 + B({\pmb k})^2 + A({\pmb k}) C({\pmb k}) + B({\pmb k}) D({\pmb k})\right] \ .
\end{eqnarray}
These functions are easily updated for each new configuration in a Monte Carlo simulation. The electrostatic energy coming from the short-range part of the potential is $U_S=(1/2)\sum_{i=1}^{N}q_i\phi_i^S({\pmb r}_i)$, where $\phi_i^S({\pmb r})$ is given by the Eq.~\ref{phi_S3}, and the self-energy contribution is $U_{self}=(1/2)\sum_{i=1}^{N}q_i\phi_i^{self}({\pmb r}_i)$. In the limit $x\rightarrow 0$, the $\text{erf}(x)$ function vanishes as $(2/\sqrt{\pi})x$ and the self-energy contribution reduces to, $U_{self}=(\kappa_e/\epsilon_w\sqrt{\pi})\sum_{i=1}^{N}q_i^2$.
The total electrostatic interaction energy of the ions is given by the above expressions plus the correction for the slab geometry. Yeh and Berkowitz~\cite{YeBe99} found that the regular 3D Ewald summation method with an energy correction, can reproduce the same results as the 2D Ewald method, with a significant gain in performance. 
Taking into account the dielectric discontinuity and the induced image charges,
we find the correction for the slab geometry to be
\begin{eqnarray}
U_{cor}=-\frac{\pi}{\epsilon_w V}\sum_{i=1}^{N}q_i \left[ \sum_{j=1}^{N} q_j
(z_i-z_j)^2 + \right.\nonumber \\
\left.\sum_{j=1}^{N} \gamma q_j (z_i-z'_j)^2 \right] \ ,
\end{eqnarray}
where $z'_j=-z_j$. Using the electroneutrality, this
expression can be written as
\begin{equation}\label{U_cor}
U_{cor}=\frac{2\pi}{\epsilon_w V} M_z^2 (1-\gamma) \ ,
\end{equation}
where $M_z=\sum_{i=1}^{N}q_i z_i$ is the magnetization in the $\hat{z}$ direction.

Now suppose that the system consists of  $N_c$ counterions of charge
$\alpha q$ and a wall of uniform surface charge density $-\sigma$, located at $z=0$. 
We first derive the
functions $A,B,C$ and $D$ appearing in the long-range part of the potential, Eq.~\ref{U_long}. For the surface charge we find
\begin{eqnarray*}
A_p({\pmb k})=-\int_{-L_{xy}/2}^{L_{xy}/2}
\int_{-L_{xy}/2}^{L_{xy}/2} \sigma\ dx\ dy \cos{(k_x x + k_y
y)}= \\
-\frac{4\sigma}{k_x k_y} \sin{(k_x L_{xy}/2)} \sin{(k_y L_{xy}/2)} \ ,
\end{eqnarray*}
\begin{equation*}
B_p({\pmb k})= \int_{-L_{xy}/2}^{L_{xy}/2}\int_{-L_{xy}/2}^{L_{xy}/2} \sigma\
dx\ dy \sin{(k_x x + k_y y)}=0 \ ,
\end{equation*}
\begin{eqnarray*}
C_p({\pmb k})=-\int_{-L_{xy}/2}^{L_{xy}/2}
\int_{-L_{xy}/2}^{L_{xy}/2}\gamma \sigma\ dx\ dy \cos{(k_x x + k_y
y)}= \\
-\frac{4\gamma\sigma}{k_x k_y} \sin{(k_x L_{xy}/2)} \sin{(k_y L_{xy}/2)}
\end{eqnarray*}
and
\begin{equation*}
D_p({\pmb k})= \int_{-L_{xy}/2}^{L_{xy}/2}\int_{-L_{xy}/2}^{L_{xy}/2}
\gamma\sigma\ dx\ dy \sin{(k_x x + k_y y)}=0 \ .
\end{equation*}
The corresponding functions for $N_c$ counterions and the charged wall 
are then: $A({\pmb k})=
\alpha q\sum_{i=1}^{N_c} \cos{({\pmb k}\cdot {\pmb
r}_i)} + A_p({\pmb k})$, $B({\pmb k})=- \alpha q\sum_{i=1}^{N_c} \sin{({\pmb
k}\cdot {\pmb r}_i)} $,
$C({\pmb k})= \gamma \alpha q\sum_{i=1}^{N_c} \cos{({\pmb k}\cdot {\pmb r}'_i)} +
C_p({\pmb k})$ and $D({\pmb k})=- \gamma \alpha q\sum_{i=1}^{N_c} \sin{({\pmb k}\cdot
{\pmb r}'_i)} $, and the total long-range part of the energy, $U_L$, is given by the Eq.~\ref{U_long}.

The short-range contribution to the electrostatic potential created by the charged surface at distance $z_i$ is
\begin{eqnarray}
\phi_p(z_i)=-\frac{2 \sigma}{(\epsilon_c+\epsilon_w)}\times \nonumber \\
\int_{-L_{xy}/2}^{L_{xy}/2} dx \int_{-L_{xy}/2}^{L_{xy}/2} dy\
\frac{\text{erfc}(\kappa_e\sqrt{x^2+y^2+z_i^2})}{\sqrt{x^2+y^2+z_i^2}} \ .
\end{eqnarray}
The limits of integration are defined in order to keep the minimum image
convention. 
We calculate the potential
on a grid in the $\hat{z}$ direction with spacing between the points 
$0.01$~\AA.  The calculation is performed once at
the beginning of the simulation, and the potential is tabulated. 
The total short range
electrostatic interaction energy is then given by $U_S=(\alpha q/2)\sum_{i=1}^{N_c}\phi_i^S({\pmb r}_i) +
\alpha q\sum_{i=1}^{N_c} \phi_p(z_i) $, where $\phi_i^S({\pmb r})$ is 
\begin{equation}
\phi_i^S({\pmb r})=\alpha q\sum_{j=1}^{N_c}{}^{'} \frac{\text{erfc}{(\kappa_e |{\pmb
r}-{\pmb r}_j|)}}{\epsilon_w |{\pmb r}-{\pmb r}_j|} + \gamma \alpha q\sum_{j=1}^{N_c}
\frac{\text{erfc}{(\kappa_e |{\pmb r}-{\pmb r}'_j|)}}{\epsilon_w |{\pmb r}-{\pmb
r}'_j|} \ .
\end{equation}
The self energy can be written as $U_{self}=(\kappa_e/\epsilon_w\sqrt{\pi})( N_c
\alpha^2 q^2 + \sigma^2 L_{xy}^4)$.  Since the charged surface is located at $z=0$, it does not contribute to the 
correction potential, Eq.~\ref{U_cor}, so that the magnetization remains 
$M_z=\alpha q\sum_{i=1}^{N_c}z_i$. The total energy
used in the simulations is
\begin{equation}
 U=U_S+U_L-U_{self}+U_{cor} . 
\end{equation}

We use $1\times 10^6$ MC steps to equilibrate the
system. The configurations are saved each $100$ MC steps. The
counterionic density profiles are obtained with $80\times 10^3$ saved
uncorrelated states.

\end{appendix}

\bibliography{ref.bib} 

\begin{thebibliography}{32}
\expandafter\ifx\csname natexlab\endcsname\relax\def\natexlab#1{#1}\fi
\expandafter\ifx\csname bibnamefont\endcsname\relax
  \def\bibnamefont#1{#1}\fi
\expandafter\ifx\csname bibfnamefont\endcsname\relax
  \def\bibfnamefont#1{#1}\fi
\expandafter\ifx\csname citenamefont\endcsname\relax
  \def\citenamefont#1{#1}\fi
\expandafter\ifx\csname url\endcsname\relax
  \def\url#1{\texttt{#1}}\fi
\expandafter\ifx\csname urlprefix\endcsname\relax\def\urlprefix{URL }\fi
\providecommand{\bibinfo}[2]{#2}
\providecommand{\eprint}[2][]{\url{#2}}

\bibitem[{\citenamefont{Guldbrand et~al.}(1984)\citenamefont{Guldbrand,
  Jonsson, Wennerstrom, and Linse}}]{GuJo84}
\bibinfo{author}{\bibfnamefont{L.}~\bibnamefont{Guldbrand}},
  \bibinfo{author}{\bibfnamefont{B.}~\bibnamefont{Jonsson}},
  \bibinfo{author}{\bibfnamefont{H.}~\bibnamefont{Wennerstrom}},
  \bibnamefont{and} \bibinfo{author}{\bibfnamefont{P.}~\bibnamefont{Linse}},
  \bibinfo{journal}{J. Chem. Phys.} \textbf{\bibinfo{volume}{80}},
  \bibinfo{pages}{2221} (\bibinfo{year}{1984}).

\bibitem[{\citenamefont{Pellenq et~al.}(1997)\citenamefont{Pellenq, Caillol,
  and Delville}}]{PeCa97}
\bibinfo{author}{\bibfnamefont{R.~J.~M.} \bibnamefont{Pellenq}},
  \bibinfo{author}{\bibfnamefont{J.~M.} \bibnamefont{Caillol}},
  \bibnamefont{and} \bibinfo{author}{\bibfnamefont{A.}~\bibnamefont{Delville}},
  \bibinfo{journal}{J. Phys. Chem. B} \textbf{\bibinfo{volume}{101}},
  \bibinfo{pages}{8584} (\bibinfo{year}{1997}).

\bibitem[{\citenamefont{Levin}(2002)}]{Le02}
\bibinfo{author}{\bibfnamefont{Y.}~\bibnamefont{Levin}}, \bibinfo{journal}{Rep.
  Prog. Phys.} \textbf{\bibinfo{volume}{65}}, \bibinfo{pages}{1577}
  (\bibinfo{year}{2002}).

\bibitem[{\citenamefont{Engstrom and Wennerstrom}(1978)}]{EnWe78}
\bibinfo{author}{\bibfnamefont{S.}~\bibnamefont{Engstrom}} \bibnamefont{and}
  \bibinfo{author}{\bibfnamefont{H.}~\bibnamefont{Wennerstrom}},
  \bibinfo{journal}{J. Phys. Chem.} \textbf{\bibinfo{volume}{82}},
  \bibinfo{pages}{2711} (\bibinfo{year}{1978}).

\bibitem[{\citenamefont{Kjellander and Mitchell}(1997)}]{KjMi97}
\bibinfo{author}{\bibfnamefont{R.}~\bibnamefont{Kjellander}} \bibnamefont{and}
  \bibinfo{author}{\bibfnamefont{D.~J.} \bibnamefont{Mitchell}},
  \bibinfo{journal}{Mol. Phys.} \textbf{\bibinfo{volume}{91}},
  \bibinfo{pages}{173} (\bibinfo{year}{1997}).

\bibitem[{\citenamefont{Netz}(2001)}]{Ne01}
\bibinfo{author}{\bibfnamefont{R.~R.} \bibnamefont{Netz}},
  \bibinfo{journal}{Eur. Phys. J. E} \textbf{\bibinfo{volume}{5}},
  \bibinfo{pages}{557} (\bibinfo{year}{2001}).

\bibitem[{\citenamefont{Moreira and Netz}(2001)}]{MoNe01}
\bibinfo{author}{\bibfnamefont{A.~G.} \bibnamefont{Moreira}} \bibnamefont{and}
  \bibinfo{author}{\bibfnamefont{R.~R.} \bibnamefont{Netz}},
  \bibinfo{journal}{Phys. Rev. Lett.} \textbf{\bibinfo{volume}{87}},
  \bibinfo{pages}{078301} (\bibinfo{year}{2001}).

\bibitem[{\citenamefont{Lau and Pincus}(2002)}]{LaPi02}
\bibinfo{author}{\bibfnamefont{A.~W.~C.} \bibnamefont{Lau}} \bibnamefont{and}
  \bibinfo{author}{\bibfnamefont{P.}~\bibnamefont{Pincus}},
  \bibinfo{journal}{Phys. Rev. E} \textbf{\bibinfo{volume}{66}},
  \bibinfo{pages}{041501} (\bibinfo{year}{2002}).

\bibitem[{\citenamefont{Jho et~al.}(2007)\citenamefont{Jho, Park, Chang,
  Pincus, and Kim}}]{JhPa07}
\bibinfo{author}{\bibfnamefont{Y.~S.} \bibnamefont{Jho}},
  \bibinfo{author}{\bibfnamefont{G.}~\bibnamefont{Park}},
  \bibinfo{author}{\bibfnamefont{C.~S.} \bibnamefont{Chang}},
  \bibinfo{author}{\bibfnamefont{P.~A.} \bibnamefont{Pincus}},
  \bibnamefont{and} \bibinfo{author}{\bibfnamefont{M.}~\bibnamefont{Kim}},
  \bibinfo{journal}{Phys. Rev. E} \textbf{\bibinfo{volume}{76}},
  \bibinfo{pages}{011920} (\bibinfo{year}{2007}).

\bibitem[{\citenamefont{Abrashkin et~al.}(2007)\citenamefont{Abrashkin,
  Andelman, and Orland}}]{AbAn07}
\bibinfo{author}{\bibfnamefont{A.}~\bibnamefont{Abrashkin}},
  \bibinfo{author}{\bibfnamefont{D.}~\bibnamefont{Andelman}}, \bibnamefont{and}
  \bibinfo{author}{\bibfnamefont{H.}~\bibnamefont{Orland}},
  \bibinfo{journal}{Phys. Rev. Lett.} \textbf{\bibinfo{volume}{99}},
  \bibinfo{pages}{077801} (\bibinfo{year}{2007}).

\bibitem[{\citenamefont{Jho et~al.}(2008)\citenamefont{Jho, Kanduc, Naji,
  Podgornik, Kim, and Pincus}}]{JhKa08}
\bibinfo{author}{\bibfnamefont{Y.~S.} \bibnamefont{Jho}},
  \bibinfo{author}{\bibfnamefont{M.}~\bibnamefont{Kanduc}},
  \bibinfo{author}{\bibfnamefont{A.}~\bibnamefont{Naji}},
  \bibinfo{author}{\bibfnamefont{R.}~\bibnamefont{Podgornik}},
  \bibinfo{author}{\bibfnamefont{M.~W.} \bibnamefont{Kim}}, \bibnamefont{and}
  \bibinfo{author}{\bibfnamefont{P.~A.} \bibnamefont{Pincus}},
  \bibinfo{journal}{Phys. Rev. Lett.} \textbf{\bibinfo{volume}{101}},
  \bibinfo{pages}{188101} (\bibinfo{year}{2008}).

\bibitem[{\citenamefont{Hatlo and Lue}(2009)}]{HaLu09}
\bibinfo{author}{\bibfnamefont{M.~M.} \bibnamefont{Hatlo}} \bibnamefont{and}
  \bibinfo{author}{\bibfnamefont{L.}~\bibnamefont{Lue}}, \bibinfo{journal}{Soft
  Matter} \textbf{\bibinfo{volume}{5}}, \bibinfo{pages}{125}
  (\bibinfo{year}{2009}).

\bibitem[{\citenamefont{Hatlo and Lue}(2010)}]{HaLu10}
\bibinfo{author}{\bibfnamefont{M.~M.} \bibnamefont{Hatlo}} \bibnamefont{and}
  \bibinfo{author}{\bibfnamefont{L.}~\bibnamefont{Lue}},
  \bibinfo{journal}{Europhys. Lett.} \textbf{\bibinfo{volume}{89}},
  \bibinfo{pages}{25002} (\bibinfo{year}{2010}).

\bibitem[{\citenamefont{Samaj and Trizac}(2011{\natexlab{a}})}]{SaTr11a}
\bibinfo{author}{\bibfnamefont{L.}~\bibnamefont{Samaj}} \bibnamefont{and}
  \bibinfo{author}{\bibfnamefont{E.}~\bibnamefont{Trizac}},
  \bibinfo{journal}{Phys. Rev. E} \textbf{\bibinfo{volume}{84}},
  \bibinfo{pages}{041401} (\bibinfo{year}{2011}{\natexlab{a}}).

\bibitem[{\citenamefont{Samaj and Trizac}(2011{\natexlab{b}})}]{SaTr11b}
\bibinfo{author}{\bibfnamefont{L.}~\bibnamefont{Samaj}} \bibnamefont{and}
  \bibinfo{author}{\bibfnamefont{E.}~\bibnamefont{Trizac}},
  \bibinfo{journal}{Phys. Rev. Lett.} \textbf{\bibinfo{volume}{106}},
  \bibinfo{pages}{078301} (\bibinfo{year}{2011}{\natexlab{b}}).

\bibitem[{\citenamefont{Moreira and Netz}(2002)}]{MoNe02}
\bibinfo{author}{\bibfnamefont{A.~G.} \bibnamefont{Moreira}} \bibnamefont{and}
  \bibinfo{author}{\bibfnamefont{R.~R.} \bibnamefont{Netz}},
  \bibinfo{journal}{Eur. Phys. J. E} \textbf{\bibinfo{volume}{8}},
  \bibinfo{pages}{33} (\bibinfo{year}{2002}).

\bibitem[{\citenamefont{Wang and Ma}(2012)}]{WaMa12}
\bibinfo{author}{\bibfnamefont{Z.~Y.} \bibnamefont{Wang}} \bibnamefont{and}
  \bibinfo{author}{\bibfnamefont{Y.~Q.} \bibnamefont{Ma}}, \bibinfo{journal}{J.
  Chem. Phys.} \textbf{\bibinfo{volume}{136}}, \bibinfo{pages}{234701}
  (\bibinfo{year}{2012}).

\bibitem[{\citenamefont{Duval et~al.}(2004)\citenamefont{Duval, Leermakers, and
  {van Leeuwen}}}]{DuLe04}
\bibinfo{author}{\bibfnamefont{J.~F.~L.} \bibnamefont{Duval}},
  \bibinfo{author}{\bibfnamefont{F.~A.~M.} \bibnamefont{Leermakers}},
  \bibnamefont{and} \bibinfo{author}{\bibfnamefont{H.~P.} \bibnamefont{{van
  Leeuwen}}}, \bibinfo{journal}{Langmuir} \textbf{\bibinfo{volume}{20}},
  \bibinfo{pages}{5052} (\bibinfo{year}{2004}).

\bibitem[{\citenamefont{{dos Santos} et~al.}(2011)\citenamefont{{dos Santos},
  Bakhshandeh, and Levin}}]{DoBa11}
\bibinfo{author}{\bibfnamefont{A.~P.} \bibnamefont{{dos Santos}}},
  \bibinfo{author}{\bibfnamefont{A.}~\bibnamefont{Bakhshandeh}},
  \bibnamefont{and} \bibinfo{author}{\bibfnamefont{Y.}~\bibnamefont{Levin}},
  \bibinfo{journal}{J. Chem. Phys.} \textbf{\bibinfo{volume}{135}},
  \bibinfo{pages}{044124} (\bibinfo{year}{2011}).

\bibitem[{\citenamefont{Lue and Linse}(2011)}]{LuLi11}
\bibinfo{author}{\bibfnamefont{L.}~\bibnamefont{Lue}} \bibnamefont{and}
  \bibinfo{author}{\bibfnamefont{P.}~\bibnamefont{Linse}}, \bibinfo{journal}{J.
  Chem. Phys.} \textbf{\bibinfo{volume}{135}}, \bibinfo{pages}{224508}
  (\bibinfo{year}{2011}).

\bibitem[{\citenamefont{Gan et~al.}(2012)\citenamefont{Gan, Xing, and
  Xu}}]{GaXi12}
\bibinfo{author}{\bibfnamefont{Z.}~\bibnamefont{Gan}},
  \bibinfo{author}{\bibfnamefont{X.}~\bibnamefont{Xing}}, \bibnamefont{and}
  \bibinfo{author}{\bibfnamefont{Z.}~\bibnamefont{Xu}}, \bibinfo{journal}{J.
  Chem. Phys.} \textbf{\bibinfo{volume}{137}}, \bibinfo{pages}{034708}
  (\bibinfo{year}{2012}).

\bibitem[{\citenamefont{{Allen, M. P. and Tildesley, D. J.}}(1987)}]{AlTi87}
\bibinfo{author}{\bibnamefont{{Allen, M. P. and Tildesley, D. J.}}},
  \emph{\bibinfo{title}{Computer Simulations of Liquids}}
  (\bibinfo{publisher}{Oxford: Oxford University Press}, \bibinfo{address}{New
  York}, \bibinfo{year}{1987}).

\bibitem[{\citenamefont{Yeh and Berkowitz}(1999)}]{YeBe99}
\bibinfo{author}{\bibfnamefont{I.~C.} \bibnamefont{Yeh}} \bibnamefont{and}
  \bibinfo{author}{\bibfnamefont{M.~L.} \bibnamefont{Berkowitz}},
  \bibinfo{journal}{J. Chem. Phys.} \textbf{\bibinfo{volume}{111}},
  \bibinfo{pages}{3155} (\bibinfo{year}{1999}).

\bibitem[{\citenamefont{Levin and Flores-Mena}(2001)}]{LeFl01}
\bibinfo{author}{\bibfnamefont{Y.}~\bibnamefont{Levin}} \bibnamefont{and}
  \bibinfo{author}{\bibfnamefont{J.~E.} \bibnamefont{Flores-Mena}},
  \bibinfo{journal}{Europhys. Lett.} \textbf{\bibinfo{volume}{56}},
  \bibinfo{pages}{187} (\bibinfo{year}{2001}).

\bibitem[{\citenamefont{G\"untelberg}(1926)}]{Gu26}
\bibinfo{author}{\bibfnamefont{E.~Z.} \bibnamefont{G\"untelberg}},
  \bibinfo{journal}{Z. Phys. Chem.} \textbf{\bibinfo{volume}{123}},
  \bibinfo{pages}{199} (\bibinfo{year}{1926}).

\bibitem[{\citenamefont{Levin et~al.}(2009)\citenamefont{Levin, {dos Santos},
  and Diehl}}]{LeDo09}
\bibinfo{author}{\bibfnamefont{Y.}~\bibnamefont{Levin}},
  \bibinfo{author}{\bibfnamefont{A.~P.} \bibnamefont{{dos Santos}}},
  \bibnamefont{and} \bibinfo{author}{\bibfnamefont{A.}~\bibnamefont{Diehl}},
  \bibinfo{journal}{Phys. Rev. Lett.} \textbf{\bibinfo{volume}{103}},
  \bibinfo{pages}{257802} (\bibinfo{year}{2009}).

\bibitem[{\citenamefont{{dos Santos}
  et~al.}(2010{\natexlab{a}})\citenamefont{{dos Santos}, Diehl, and
  Levin}}]{DoDi10b}
\bibinfo{author}{\bibfnamefont{A.~P.} \bibnamefont{{dos Santos}}},
  \bibinfo{author}{\bibfnamefont{A.}~\bibnamefont{Diehl}}, \bibnamefont{and}
  \bibinfo{author}{\bibfnamefont{Y.}~\bibnamefont{Levin}},
  \bibinfo{journal}{Langmuir} \textbf{\bibinfo{volume}{26}},
  \bibinfo{pages}{10778} (\bibinfo{year}{2010}{\natexlab{a}}).

\bibitem[{\citenamefont{Shklovskii}(1999)}]{Sh99}
\bibinfo{author}{\bibfnamefont{B.~I.} \bibnamefont{Shklovskii}},
  \bibinfo{journal}{Phys. Rev. E} \textbf{\bibinfo{volume}{60}},
  \bibinfo{pages}{5802} (\bibinfo{year}{1999}).

\bibitem[{\citenamefont{Totsuji}(1975)}]{To75}
\bibinfo{author}{\bibfnamefont{H.}~\bibnamefont{Totsuji}}, \bibinfo{journal}{J.
  Phys. Soc. Jpn.} \textbf{\bibinfo{volume}{39}}, \bibinfo{pages}{253}
  (\bibinfo{year}{1975}).

\bibitem[{\citenamefont{Bakhshandeh et~al.}(2011)\citenamefont{Bakhshandeh,
  {dos Santos}, and Levin}}]{BaDo11}
\bibinfo{author}{\bibfnamefont{A.}~\bibnamefont{Bakhshandeh}},
  \bibinfo{author}{\bibfnamefont{A.~P.} \bibnamefont{{dos Santos}}},
  \bibnamefont{and} \bibinfo{author}{\bibfnamefont{Y.}~\bibnamefont{Levin}},
  \bibinfo{journal}{Phys. Rev. Lett.} \textbf{\bibinfo{volume}{107}},
  \bibinfo{pages}{107801} (\bibinfo{year}{2011}).

\bibitem[{\citenamefont{dos Santos et~al.}(2009)\citenamefont{dos Santos,
  Diehl, and Levin}}]{DoDi09}
\bibinfo{author}{\bibfnamefont{A.~P.} \bibnamefont{dos Santos}},
  \bibinfo{author}{\bibfnamefont{A.}~\bibnamefont{Diehl}}, \bibnamefont{and}
  \bibinfo{author}{\bibfnamefont{Y.}~\bibnamefont{Levin}}, \bibinfo{journal}{J.
  Chem. Phys.} \textbf{\bibinfo{volume}{130}}, \bibinfo{pages}{124110}
  (\bibinfo{year}{2009}).

\bibitem[{\citenamefont{{dos Santos}
  et~al.}(2010{\natexlab{b}})\citenamefont{{dos Santos}, Diehl, and
  Levin}}]{DoDi10a}
\bibinfo{author}{\bibfnamefont{A.~P.} \bibnamefont{{dos Santos}}},
  \bibinfo{author}{\bibfnamefont{A.}~\bibnamefont{Diehl}}, \bibnamefont{and}
  \bibinfo{author}{\bibfnamefont{Y.}~\bibnamefont{Levin}}, \bibinfo{journal}{J.
  Chem. Phys.} \textbf{\bibinfo{volume}{132}}, \bibinfo{pages}{104105}
  (\bibinfo{year}{2010}{\natexlab{b}}).

\end{thebibliography}

\end{document}